Libraries, Digital Libraries, and Data: Forty years, Four Challenges

Christine L. Borgman, Distinguished Research Professor, Information Studies, University of California, Los Angeles, CA 90095-1520; Christine.Borgman@UCLA.EDU

ABSTRACT

"Digital libraries" is an umbrella term that encompasses the automation of library services, online catalogs, information retrieval systems, multi-media databases, data archives, and other internet-facing collections of digital resources. Clifford Lynch has played pivotal roles in the technical development, institutionalization, policy, practice, and dissemination of digital libraries for more than 40 years. Beginning with his foundational role in building MELVYL for the University of California in the early 1980s – the first internet-native online open access library catalog – through his convening roles in open access and open data in the 21$^{st}$ century, his career is marked by multiple milestones of innovation. Clifford Lynch's career has traced the trajectory of digital libraries and knowledge infrastructures. Over the course of these 40 years, research libraries have faced four categories of challenges: invisible infrastructure, content and collections, preservation and access, and institutional boundaries. These challenges have become yet more complex in an era of open access, open data, and evolving regimes of intellectual property and scholarly publishing. As the digital library communities have merged and diverged over this time span, we collectively face challenges for at least the next 40 years to sustain access to current resources while growing the next generations of digital libraries and librarians.

## The Dawn of Online Catalogs

Clifford Lynch began his pioneering career in research libraries when he joined the University of California Division of Library Automation (UC-DLA) in 1980 to lead the development of MELVYL, the first major online library catalog designed for Internet access (Ashenfelder, 2013). In these early days, universities were constructing their own automated library systems for access via campus networks or dial-up modem. The nascent market for automated library systems did not yet scale up to the needs of large, distributed institutions, whether university or public libraries. Following a path parallel to Clifford Lynch, the author began her career in library automation as Systems Analyst for the Dallas Public Library, with primary responsibility for developing an online catalog. Written in assembler language on the City of Dallas mainframe computer system, the catalog went live for patron access in late 1978 on CRT (cathode ray tube) terminals (Borgman, 1977; Borgman & Kaske, 1980).

Also in 1980, the Council on Library Resources[1] (CLR) funded the first major study of online public access catalogs[2]. Partners in the CLR collaboration included UC-DLA, Research Libraries Group (RLG)[3], OCLC[4], and the Library of Congress (McCarn, 1983; Research Libraries Group, 1982). Clifford was the delegate for the University of California. I was a delegate for RLG, based at Stanford, where I was a doctoral student and research assistant on the OPAC project. Later I represented OCLC in several OPAC studies (Borgman, 1983, 1986a). The CLR study laid the groundwork for technology, governance, and usability of automated library systems as the market and technologies began to mature (Kochtanek & Matthews, 2002; Matthews, 1982). Our collaboration on the CLR study led to a 40+ year conversation with Clifford Lynch about libraries, information technologies, infrastructures, institutions, policies, and much else.

**Emerging Knowledge Infrastructures**

Only in the early 1990s, as the Cold War ended, and as Western countries lifted their technology blockades to the Soviet Bloc, did the Internet become a functioning international infrastructure with open interconnections across countries and sectors. MELVYL, as the first Internet-facing online library catalog, and probably the largest, including resources from then nine campuses[5] of the University of California (UC), became an international resource. During my years of teaching library automation and information retrieval at UCLA (1983-), and as a Fulbright Professor in Budapest (1993), MELVYL was a primary instructional resource. The UC-wide catalog offered far more resources and search capabilities than anything else publicly available at the time, especially in Hungary or Central and Eastern Europe.

While Al Gore did not 'invent the Internet[6],' he did play important roles in information policy that enabled international network interconnections (Kronenfeld & Kronenfeld, 2020). At the Inaugural World Telecommunication Development Conference in 1994, the new policy was announced as the *Global Information Infrastructure (GII)*(Gore, 1994). As the GII became a functioning infrastructure, the library world was well positioned to take advantage (Borgman, 2000a). The vision of the GII as a 'global digital library' built upon investments by the library community in technology, standards, institutions, and infrastructure such as the MARC format, Text Encoding Initiative, Dublin Core metadata, Unicode, OCLC, and RLG. National libraries also played important roles in developing information infrastructure within and between countries. Clifford Lynch was on the front lines of the GII, writing about topics such as research integrity, persistent identifiers, search technologies, and digitization (Lynch, 1992, 1994, 1997, 1998a, 2000a; Lynch & Brownrigg, 1986; Tyner & Lynch, 1980), all the while advancing the digital library services of the University of California and pursuing his doctorate in computer science at Berkeley.

---

[1] The Council on Library Resources (CLR) later became the Council on Library and Information Resources (CLIR).
[2] The terms Online Public Access Catalog, and associated acronym, 'OPAC,' were coined by the CLR study participants.
[3] The Research Library Group later was acquired by OCLC and their databases were merged.
[4] Then known as the Ohio College Library Center.
[5] UC Merced, the tenth campus, opened in 2005.
[6] An urban legend, oft refuted.

As terms such as 'information infrastructure' came into common usage, scholars began to explore the historical origins of infrastructure in railroads, radio, telecommunications, finance, and other sectors (Brown et al., 1995; David, 2005; Edwards, 2003; Friedlander, 1995a, 1995b, 1996a, 1996b, 2005; Joint Information Systems Committee, 2009; Ribes & Finholt, 2009; Ribes & Lee, 2010). Infrastructures are comprised of interacting technical, institutional, policy, and social components that evolve over time. Star and Ruhleder (1996), taking a socio-technical perspective, identified eight dimensions of infrastructure, a model commonly recognized in the information science community: infrastructures are *embedded* in other social arrangements and technologies; *transparent*, supporting tasks invisibly; are *spatial or temporal* in scope; *learned as part of membership* in an organization; *linked with conventions of practice* in regular work; an *embodiment of standards*; *build upon an installed base*, inheriting capacity and limitations; and become *visible upon breakdown*, when they cease to function as expected. Concerns for libraries, digital libraries, and data are now subsumed in the context of knowledge infrastructures – defined as 'robust networks of people, artifacts, and institutions that generate, share, and maintain specific knowledge about the human and natural worlds' (Edwards, 2010, p. 17).

The mid-1990s were optimistic times for library technologies, but online public access catalogs (OPACs) were fledgling systems that remained difficult to use (Borgman, 1986b, 1996). In response to a groundswell of interest in advancing information retrieval technologies, a coalition of ten US funding agencies, led by the National Science Foundation, launched a ten-year (1994-2004) multi-disciplinary funding program known as the Digital Libraries Initiative (DLI) (*Digital Libraries*, 1999; Griffin, 1998). UC-Berkeley was among the funding recipients and again Clifford Lynch played a role. Of the countless number of papers resulting from 10 years of the DLI, the best known is the PageRank algorithm introduced as Google (Brin & Page, 1998), initially a poster presentation at a DLI All-Hands Meeting at Stanford University. Similarly influential within the digital libraries community was the research agenda that established basic principles for interoperability and infrastructure (Lynch & Garcia-Molina, 1995). Clifford Lynch and Hector Garcia-Molina[7] later revisited these interoperability issues as technology evolved (Hey et al., 2006; Lynch & Garcia-Molina, 2002).

It was during the Digital Libraries Initiative that Paul Evan Peters, founding director of the *Coalition for Networked Information*, died suddenly. Clifford was the only imaginable candidate with the breadth of knowledge and deep engagement in the community to step into the CNI role in 1997. We in the University of California were sorry to lose him, but he remains close to the community[8]. As of this writing, Clifford Lynch and Michael Buckland continue to offer a Friday afternoon seminar at UC-Berkeley on information systems.

---

[7] Sergei Brin and Larry Page, who created Google as a DLI project, were doctoral students in computer science at Stanford University under the supervision of Hector Garcia-Molina and Terry Winograd.

**Diverging Communities**

While the digital libraries initiative was a coalition of libraries and computer science researchers, the focus of these communities began to diverge fairly quickly. By the latter 1990s, computer scientists were framing digital libraries as advanced information retrieval systems that supported content in multiple media. Libraries, in contrast, viewed digital libraries more holistically in terms of serving the information needs and uses of user communities (Borgman, 1999).

The latter 1990s, in Clifford Lynch's early years as director of CNI, were marked by a series of existential challenges to research libraries. In the 1970s era of library automation, computer scientists too often asked, "why does the library need a computer?". Twenty years later, the next generation of computer scientists – and some university leaders – were asking the inverse question, "now that we have computers, why do we need libraries?". Clifford addressed these challenges head-on in a series of papers aimed at library, technology, academic, and policy audiences (Lynch, 1997, 2000a, 2000b, 2001a, 2001b, 2002a, 2002b, 2003a, 2003b, 2003c, 2004). In these and other talks and papers, he was among the first to address questions of authenticity, integrity, and digital stewardship as concerns of policy and practice. Assessing the past and future of digital libraries, Lynch (2005) commented that the term 'digital library' had become an oxymoron, disconnected from libraries as institutions or the practice of librarianship.

Lynch's public talks were already legendary, anticipating technology and policy trends several years out. Also legendary was his travel schedule, flying around the globe to attend conferences and meetings on countless topics related to networked information. Clifford is the ultimate networked individual who transports news across community boundaries. No amount of reading about current developments in information technology and policy could result in as much knowledge acquisition as listening to one of his talks or better yet, a long after-hours discussion. He is among the few speakers who can present a coherent framework over the course of an hour, from a few bullet points jotted on a notecard. Until video recording became common practice, few artifacts – and no slide decks – of those influential talks remain.

In the three decades since the launch of the Digital Libraries Initiative, research on digital libraries has expanded across domains of knowledge and communities, incorporating content in more media, both digitized and born digital. The *Joint Conference on Digital Libraries* series launched in 2001 as a partnership of two computer science societies, the *Association for Computing Machinery* and *IEEE*, with the explicit goal of bringing library and computer science communities together (Borgman & Hessel, 2001; Fox & Borgman, 2001). While the initial conferences were successful convening events, later efforts diverged, with JCDL becoming dominated by computer science and engineering research, where much more funding for research and travel is available. JCDL continues to attract international participants and holds conferences in the U.S. and abroad. It was joined by the *European Digital Libraries Conference* (later becoming *Theory and Practice of Digital Libraries*), the *Asian Conference on Digital Libraries*, and numerous smaller conferences and workshops that bridge communities.

---

[8] Clifford Lynch frequently visited UCLA to guest-lecture my library technology courses and to meet with others on campus.

**Four Challenges for Libraries**

While the growth of research and publishing in digital libraries advanced the technological frontier for multiple economic sectors, public concern focused largely on the *digital* rather than on *libraries*. The library community was decreasingly visible in conferences and journals about digital libraries. As we entered the 21$^{st}$ century, four challenges for libraries in a digital age came into view: invisible infrastructure, content and collections, preservation and access, and institutional boundaries (Borgman, 2000a, 2000b, 2003). None of these challenges are yet resolved, 25 years on. Rather, they have changed in form and scale over time, becoming ever more critical for the future of libraries and librarianship.

*Invisible infrastructure*

By the year 2000, a troubling refrain was emerging, perhaps common to all except humanities scholars, that people no longer visit 'the library' because everything they need is online (Borgman, 2003). A defining characteristic of infrastructure is that is invisible until it breaks down (Star & Ruhleder, 1996). Invisibility is a sign of success, in that libraries were serving their communities so effectively that their existence was receding into the background. Invisibility is also a sign of failure, in that the substantial expertise, labor, and resources necessary to provide those services goes unrecognized. In times of shrinking budgets, invisibility is especially dangerous.

The work of Clifford Lynch and CNI has been essential to maintaining the visibility of library systems and services in the broader educational and policy communities. Access to information is seamless only to the extent that systems are interoperable, which requires substantial investments in technical standards, and deep expertise in knowledge organization. Technologies and partnerships that developed in the first two decades of the 21$^{st}$ century, such as Digital Object Identifiers (DOI), ORCID, and Crossref, are the foundation for today's access to bibliographic content. They have become the invisible plumbing of knowledge infrastructures to those outside the library and information technology community.

*Content and collections*

The rise of digital libraries put the notion of library collections into sharp relief (Buckland, 1988, 1992). Research libraries were less able to compete for status based on the sheer number of volumes they acquired. Access to content, whether owned, licensed, or linked, became a growing concern. Here again, Clifford Lynch helped to lay the intellectual groundwork for maintaining the coherence of library collections, whether physical or digital, local or remote (Lynch, 1999). As the library and computer science communities diverged, so did concerns for coherent knowledge organization. Libraries catalog materials they acquire to support the information needs of their communities. They have far less control over how remotely accessed content is organized. Library catalogs increasingly are subsumed under content management systems, and users find their way into many of these resources via search engines external to library control.

*Preservation and access*

Despite the lack of agreement on what constitutes a library 'collection,' the need to preserve collections in ways that they remain accessible became a growing challenge. Here again, Clifford waded through the sea of definitions of 'preservation' and 'access' to offer a path forward (Lynch, 1998b, 2002a, 2003d). He called attention to the crises ahead as physical materials were crumbling and as publishers controlled access to digital content, without assurances of sustaining the viability of those resources. As a member of several study panels convened by the U.S. National Academies, Lynch made sure that concerns of the research library community were represented (Berman et al., 2010; *LC21*, 2000; *Life-Cycle Decisions for Biomedical Data*, 2020). The lack of coherent economic and policy models for preserving and sustaining access to digital collections remains a massive problem for libraries – past, present, and future.

*Institutional boundaries*

Collectively known as 'memory institutions,' libraries, museums, and archives are tasked with organizing, preserving, and providing access to knowledge in various forms. Despite their common interests, each of these institutions has distinct histories, practices, and theoretical frameworks for collecting and organizing knowledge. Digitization has blurred the boundaries between types of artifacts, such as books, records, and objects, and between institutions. Determining who collects what kinds of content, and sustains access for future generations, becomes even more complex when publishers lease access to content, and as research data repositories are hosted by funding agencies or as independent entities. The CNI community has broadened their tent over the last several decades to include more stakeholders from archives, museums, data repositories, publishing, technology, and policy to negotiate boundaries and partnerships (Committee on Future Career Opportunities and Educational Requirements for Digital Curation et al., 2015; Kennedy & Lynch, 2020; Lynch, 2000b, 2003c, 2006, 2013, 2017).

**Open Access, Open Data**

Electronic publishing began in earnest in the 1990s and became the norm by the early 21$^{st}$ century, contributing to massive changes in the publishing industry and in library services (Borgman, 2007). Scholars initially lacked trust in online journals, leading to hybrid journals published in both print and digital form, and to complex debates over which edition constitutes 'the version of record.' By 2001, Clifford Lynch (2001a) was exploring 'the future of the book' as a metaphor for rethinking the roles of cultural artifacts and intellectual property in an increasingly digital world.

The apparent ease of digital distribution obscured the labor and costs involved in scholarly publishing, leading to claims that 'information wants to be free'… and expensive[9] (Wagner, 2003). Anarchy in the early days of digital publishing gradually gave way to massive restructuring of the publishing industry and to many flavors of open access distribution. Setting aside complex formulae of who pays for what, when, and how, a simple definition of 'open access' is that content is free to the reader. Various colors of open access are now the norm in scholarly publishing, promoted by governments and funding agencies in much of the world.

---

[9] Attributed to Stewart Brand by multiple authors.

Open access and related developments in scholarly communication brought opportunities to research libraries, but also wrought restructuring across the four challenges identified above (Borgman, 2007). Library investments in infrastructure continued to expand as collections diversified in format, content, and contract. Sustaining access to materials not owned by library institutions was a growing problem. Memory institutions began to work more closely across boundaries on matters of knowledge organization and access. Again, Clifford Lynch and CNI gave voice to library perspectives on digital preservation, institutional repositories, economics of digitization, misinformation, and university roles in disseminating scholarship (Lynch, 2001b, 2002a, 2003a, 2003b, 2003d, 2004; Lynch & Lippincott, 2005).

Educational implications of knowledge infrastructures were another growing concern in the first decade of the 21$^{st}$ century. CNI provides a bridge between research libraries and university infrastructure as a partnership between the Association for Research Libraries and EDUCAUSE. This partnership was leveraged to good effect by addressing new models for learning and for open educational resources made possible by new infrastructure investments (Borgman et al., 2008; Lynch, 2002b, 2003c).

Roughly parallel in time with the growth of open access publishing, if somewhat slower to reach critical mass, was growth in open data policies. Open data, in the simplest definition, is releasing datasets associated with scholarly publications (Borgman, 2015; Pasquetto et al., 2016). Digital datasets are an unusual category of materials that rarely are collected by libraries, archives, or museums. Discipline-specific data archives, where they exist, are the more common home for such resources. Data formats, content, and scale vary by orders of magnitude from astronomy to archaeology, and for everything in between. CNI and Lynch addressed these challenges to universities early on, articulating characteristics of, and requirements for, stewarding datasets ("Big Data in the Campus Landscape," 2014; Committee on Future Career Opportunities and Educational Requirements for Digital Curation et al., 2015; Lynch, 2008, 2009, 2013, 2014, 2017).

Managing research data is a growing priority for the scholarly enterprise, leading to broader concerns for research data infrastructure (Borgman, 2023; Directorate-General for Research and Innovation (European Commission) et al., 2021). Funding agency requirements for sharing research data are based on solid arguments for leveraging investments in research, promoting transparency, and reproducibility of findings (Borgman & Bourne, 2022; National Academies of Sciences, 2019, 2021). However, individual researchers often have few incentives to share their data. They also have valid concerns about misuse, misinterpretation, and misappropriation of their work. Making research data FAIR (findable, accessible, interoperable, reusable) remains complex, labor-intensive, and expensive in most disciplines (Borgman & Groth, 2025; Wilkinson et al., 2016). Making research data sustainable for the long-term involves an array of economic, social, institutional, and infrastructural challenges. The lack of sufficient data management workforce comprised of data librarians and archivists is central to these challenges.

**The Next 40 Years**

Online public access catalogs were known initially as 'online card catalogs,' reflecting the technological transition under way. Few of today's students grew up with card catalogs, making card-based user interfaces appear antiquated. Librarians have invested more than 40 years of labor in adapting standards and practices for knowledge organization to a world of digital content and networked institutions (Kennedy & Lynch, 2020; Sanson, 2024; *UNIMARC Committee*, 2024; Willer & Dunsire, 2013; Willer & Katić, 2024). An existential challenge for libraries is to maintain continuity with centuries of cataloging practice while developing modern practices that interoperate with technical standards developed (and controlled) by other agencies, such as the World Wide Web Consortium.

Today's generation of 'digital natives' bring their knowledge of search engines and web-based interfaces to library catalogs and databases. 'User training' in online searching is a foregone task (Borgman, 1981). However, keyword searching alone is insufficient to exploit the full power of library cataloging systems that employ nuanced distinctions between editions, formats, languages, similar author names, and other bibliographic details. Library users who employ skills learned on web search engines rarely know what they are missing unless they consult librarian experts. A larger concern is that many library users assume that everything they need exists online, in digital form. Vast amounts of valuable scholarly material exist only in physical form, large portions of which are minimally described or remain uncataloged. Hidden collections abound. When cultural collections are lost due to fires, flood, climate change, war, or other disasters, the world may lack records of what existed. More than 20 years ago, Clifford Lynch warned us of the coming crisis in sustaining access to these resources, proposing digitization at least as backup against disaster (Lynch, 2003d).

A looming challenge for the next 40 years is how to sustain access to the vast resources of digital and digitized content now held by libraries, museums, archives, governments, business, and individuals. Most physical artifacts such as books and journals can survive by benign neglect. Digital resources survive only by continual refreshment and migration to new technologies. Research datasets are especially fragile because they often rely on custom software tools and platforms that evolve rapidly (Borgman & Groth, 2025; Goodman et al., 2014). Underlying these infrastructural challenges for sustaining access to scholarship are human resource constraints. The necessary workforce of skilled librarians, data managers, data archivists, and data curators is sorely lacking in universities, research domains, and in other sectors of the economy. This problem is not new, nor are the reasons for the lack of workforce investment simple (Committee on Future Career Opportunities and Educational Requirements for Digital Curation et al., 2015; Dobozy, 2013; Lane, 2024; Marshall et al., 2009; *Research Skills for an Innovative Future*, 2011; "Workforce Issues in Library and Information Science, Part 2," 2010; "Workforce Issues in Library and Information Science: Special Issue," 2009; Weber et al., 2012).

The better the technology and infrastructure, the less inherent is visibility of the investments made by libraries and librarians. More than ever, CNI and other library leaders must make visible the invisible roles of their institutions.

**Acknowledgements**
The historical framing in this paper is based loosely on my keynote presentation to the 2024 Conference, *Theory and Practice of Digital Libraries*, in Ljubljana, Slovenia (Borgman, 2024); thanks are due to Gianmaria Silvello, Ines Vodopivec, and the National and University Library of Slovenia for the invitation and conference hosting. I cherish my 45-year friendship with Clifford Lynch, his many insights in our long and frequent conversations, and his generous editing and advising on all three of my MIT Press books cited herein (Borgman, 2000a, 2007, 2015).